# FLASH Beam-Off RF Measurements and Analyses

Shilun Pei[1], Chris Adolphsen[1] and John Carwardine[2] [*]

1 – SLAC National Accelerator Laboratory
2575 Sand Hill Road, Menlo Park, CA 94025– United States

2 – Argonne National Laboratory – APS ASD Division
9700 S. Cass Avenue, Argonne, IL 60439 – United States

The FLASH L-band (1.3 GHz) superconducting accelerator facility at DESY has a Low Level RF (LLRF) system that is similar to that envisioned for ILC. This system has extensive monitoring capability and was used to gather performance data relevant to ILC. In particular, waveform data were recorded with beam off for three, 8-cavity cryomodules to evaluate the input rf stability, perturbations to the SC cavity frequencies and the rf overhead required to achieve constant gradient during the 800 μs pulses. In this paper, we discuss the measurements and data analysis procedures and present key findings on the pulse-to-pulse input rf and cavity field stability.

## 1 Introduction

The FLASH (Free-Electron LASer in Hamburg) facility at DESY is the world's only free-electron laser for VUV and soft X-ray production. Its layout is shown in Figure 1. The electron bunches are produced in a laser-driven photo-injector and accelerated by a superconducting linear accelerator. At intermediate energies of 127 MeV and 450 MeV the 1 nC bunches are longitudinally compressed from several picoseconds to several tens of femto-seconds, thereby increasing the peak current from 50-80 A to approximately 1-2 kA as required for FEL operation.

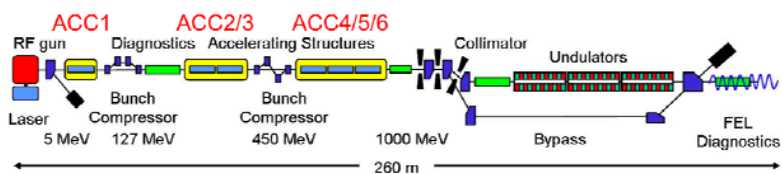

Figure 1: Layout of the FLASH facility.

Presently there are six accelerator modules each containing eight, L-band (1.3 GHz), 1-m long, 9-cell, superconducting cavities. The three modules, ACC4-ACC6, are the focuses of this study as they are very similar to an ILC rf unit. These 24 cavities are powered by a single klystron and the LLRF system monitors the input and reflected rf at each cavity as well as the cavity fields using probe couplers. The probe signals for the 24 cavities are summed vectorally and used in Feed Back (FB) and Adaptive Feed Forward (AFF) algorithms to keep the net gradient from the 24 cavities constant during the 800 μs beam period that follows a 500 μs cavity fill period. These algorithms control the drive rf to the klystron in this process. The AFF corrections try to incorporate the repetitive pulse-to-pulse corrections made by the FB system.

LLRF waveform data were collected on 09/18/08 with the beam off and without cavity piezo actuator compensation for Lorentz force detuning. Three sets of the data were taken in which: 1) FB and AFF were off; 2) FB was on and AFF was off; 3) FB and AFF were on. For each set, the phase and amplitude waveforms for the rf input, reflected and probe signals of all 24 cavities were recorded simultaneously for 100 pulses using the DOOCS control system. Here the pulses could only be measured at a rate of about 1/3 Hz so the time to acquire one data set was ~ 3 seconds [2]. In the future, we plan to use the FLASH DAQ archiver system [3] to increase the data collection frequency to 5 Hz, which is the machine repetition rate.

[*] Work supported by the DOE under contract DE-AC02-76SF00515.

The main purpose of this study was to measure the input and cavity rf stability, the latter of which is affected by Lorentz force detuning and microphonic induced cavity frequency changes. For this study, the FB and AFF off data are mainly relevant. We also wanted to determine the input rf overhead requirement when piezo actuators are used to compensate the main effect of Lorentz force detuning, as would be the case in the ILC. However, this actuator system is not yet automated at FLASH and the data were taken without this compensation. In this case, the rf overheads observed with the FB and/or AFF on are larger than would be with such compensation. Only beam-off data were recorded as the FLASH beam differs significantly in both intensity and stability compared to that expected at ILC.

## 2  Data analysis

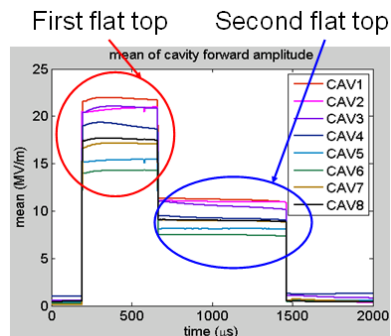

Figure 2: Typical rf input waveforms for cavities in ACC6 with FB and AFF off.

For superconducting cavities operated without beam, the input rf pulse has two levels (i.e., two flattop regions) that correspond to the cavity fill period and the nominal constant gradient period (see Figure 2). To maintain the cavity gradient with no beam loading, the rf input amplitude is halved during the second period.

For the vector sum signals (i.e., the vector sum of the 24 calibrated cavity probe signals) and the cavity input and reflected signals, a time domain analysis was done where the mean and standard deviation of the 100 pulses was computed versus time during the pulse. For the individual cavity probe signals, this analysis was also performed, and in addition, a frequency domain study was done for the 100 pulses at two selected times during the pulse.

### 2.1  Time Domain Analysis

The time domain analysis for the cavity input signals can be divided into four basic steps. In the first step, the cavity input signals were calibrated in units of MV/m using the cavity probe signals as a reference. The probe signals had been calibrated prior to the data taking but the rf input signals had not. Linking the rf field during the first flattop to the cavity gradient provides fairly accurate relative calibration since the cavity $Q_{ext}$ values are nominally all the same (3e6). The second step involved averaging each set of four data points. That is, the data were sampled at 1 MHz and this procedure eliminates any 250 kHz LLRF reference signal leakage [4]. For the third step, the standard deviation and mean value at each data sample time for the 100 pulses were computed for each set of measurements. The standard deviation before the rf turns on is a measure of electronic noise, while that when the rf is on includes both the rf jitter and electronic noise. So in the fourth step, the electronic noise contribution was subtracted in quadrature from the input rf signals based on the rf-off baseline value. Since relative effects are of most interest, the ratio of the jitter (standard deviation) to the mean amplitudes were computed, which we call Percentage Standard Deviation (PSD).

For the time domain analysis of the vector sum signals, both the PSD of the amplitude and the ASD (Absolute Standard Deviation) of phase were computed, but the relatively small electronic noise contribution to the amplitude PSD was not subtracted. For the cavity reflected and probe signals, the mean value and PSD including electronic noise were computed, respectively.

### 2.2  Frequency Domain Analysis

The frequency domain method used to analyze the cavity probe signals is illustrated in Figure 3. In the "Sample" space, there are 2048 data points (1 µs/point) for each "Pulse" number; while in "Pulse" space, there are 100 data points (~ 3 sec/point) at each "Sample" time. The frequency domain analyses were done in the "Pulse" space at two fixed "Sample" times: the beginning and end of the flattop for the cavity probe signals.

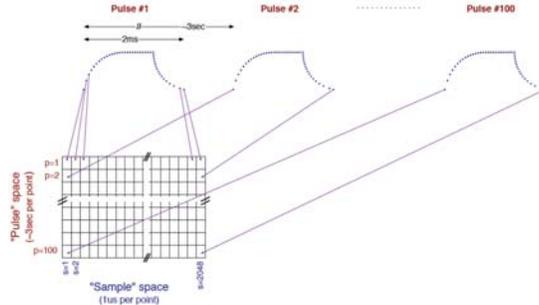

Figure 3: 'Pulse' and 'Sample' space illustration.

## 3  Key Findings

Our analyses show that the FB and AFF algorithms work well to reduce the jitter and flatten the vector sum amplitude and phase. Also, the input rf signals are very stable (~ 0.1% amplitude jitter) with both FB and AFF off, indicating that the klystron modulator and rf drive systems have very small pulse-to-pulse variations. The cavity probe signal jitter is dominated by variations in the pulse-to-pulse cavity detuning; the jitter is essentially random pulse-to-pulse with large cavity-to-cavity variations that are not significantly correlated among cavities. In addition, the cavity field jitter doesn't scale with the square of cavity gradient in 14-24 MV/m range as would be expected for Lorentz force detuning. This suggests there may be large variations in the mechanical stiffness among the 24 cavities (a factor of two might be expected) and/or there are local sources driving the cavity vibrations. More details on these findings are presented in the sections below.

### 3.1  Vector Sum Signals

Figure 4 and 5 show the vector sum amplitude and phase data, respectively. With both FB

and AFF on, the flattop PSD decreases from 0.1-0.4% to 0.05-0.08%, the phase ASD decreases from $0.25$-$0.45^o$ to $0.06$-$0.075^o$. The FB works very well to reduce the amplitude PSD and phase ASD, but it is not enough to flatten the pulse. Fortunately, AFF can flatten the pulse to the resolution limit.

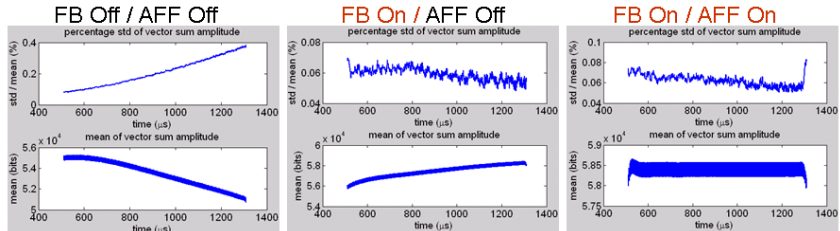

Figure 4: PSD and mean vector sum during the flattop period.

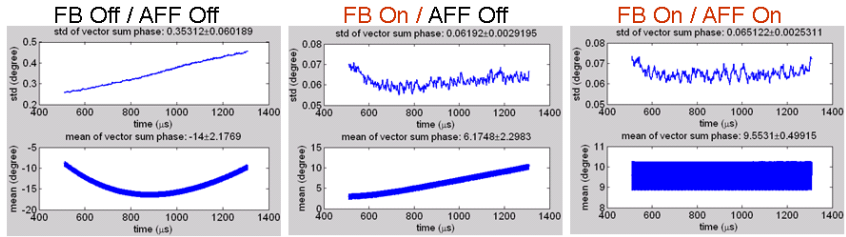

Figure 5: ASD and mean vector sum phase during the flattop period.

### 3.2 Cavity Input and Reflected Signals

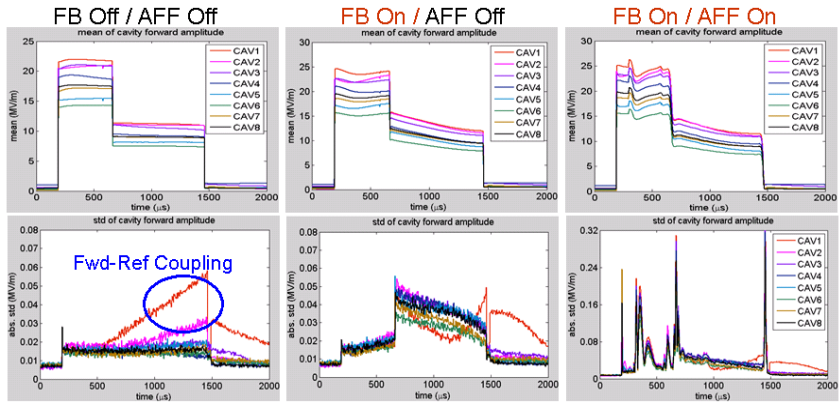

Figure 6: Mean and ASD of the input (forward) rf signals for cavities in ACC6

Figure 6 shows the mean and ASD of the calibrated input rf signals for cavities in ACC6. The signals are fairly noisy compared with the input rf jitter, and there is significant reflected-to-input signal coupling, making it difficult to de-convolute these signals. That is, the input rf signal as measured is proportional to the sum of actual input rf plus some fraction of the reflected rf, which is fairly large during the fill period. The input-to-reflected isolation of the directional couplers used to measure the signals is ~ 20 dB whereas it would have been better if DESY uses ones with at least ~ 35 dB isolation. This coupling causes slight differences in the input rf flattop shapes for the different cavities as can be seen in Figure 7 (here the waveforms are normalized to the same average amplitude to more easily discern the differences - the actual input rf pulse shapes should be identical as the rf is generated by one klystron). The jitter on the input signal is also strongly affected by this coupling in some cases. For example, the 'CAV 1' input jitter (shown in red in the bottom left plot in Figure 6) is dominated by the jitter on the reflected rf that couples into the input rf measurement.

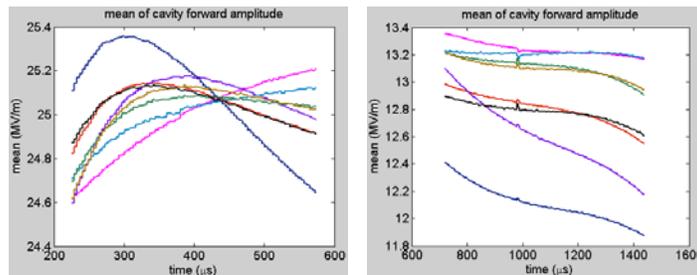

Figure 7: Normalized input (forward) rf waveforms for cavities in ACC6 during the first (left) and second (right) flattop periods.

Ideally, the reflected rf at the end of the first flattop period should be zero. That is, given the cavity $Q_{ext}$, the fill time is set so the reflected power goes to zero at the end of fill period if the cavity is running on-frequency. Instead, one sees that the reflected rf amplitude at the end of the first flattop is around 50% of initial reflection during the first flattop (the initial reflection roughly equals the input rf amplitude). Figure 8 is plot of this fraction for all cavities. This large ratio suggests that the cavities are running fairly far off resonance and have significant initial detuning. This is not surprising as Lorentz force detuning during the fill period for 20 MV/m operation is expected to be roughly half of the cavity bandwidth [4].

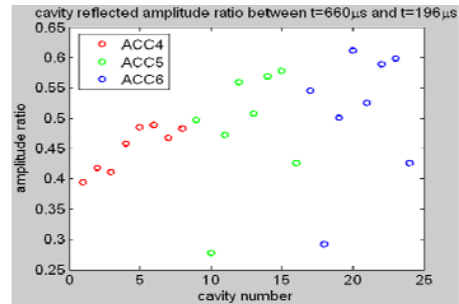

Figure 8: Ratio of reflected rf at the end of the first flattop to the initial reflection during the first flattop.

Figure 9 and Figure 10 show the input rf PSD for all cavities averaged over the two flattop regions with the noise contribution subtracted in quadrature. The error bars on the points encompass the range of jitter during the flattop periods. With FB and AFF off, the flattop amplitude is very stable pulse to pulse; the fractional jitter is ~ 0.07% for first flattop and 0.15% for the second. This factor of two increase suggests the jitter originates from noise in the rf drive as the absolute rf jitter is independent of amplitude. This differs from usual case in which modulator voltage variations generate proportional rf jitter. The cavities with high PSD are ones where there is a large jitter on the reflected signal and the reflected-to-input coupling dominates the actual input rf jitter.

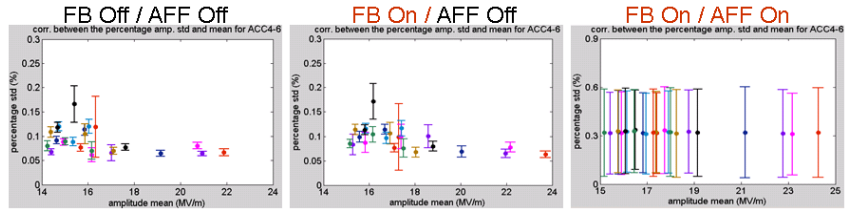

Figure 9: Input signal PSD averaged over the first flattop for all cavities.

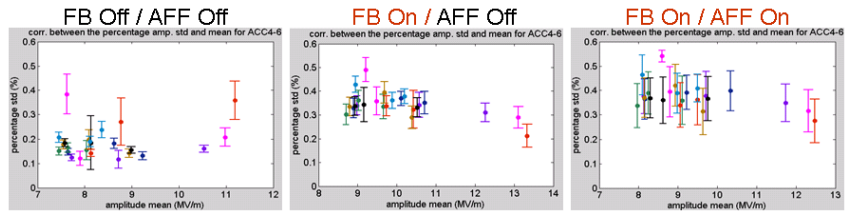

Figure 10: Input signal PSD averaged over the second flattop for all cavities.

With FB on and AFF off, the cavity input waveform shape changes as expected for Lorentz force detuning without piezo compensation. With both FB and AFF on, rapid corrections were added to the input waveform during the first flattop, and are the main cause of jitter. It would be better to apply the AFF corrections more smoothly during the first flattop to reduce the required rf overhead.

### 3.3 Cavity Probe Signals

Figure 11 shows the cavity probe signals of selected pulses during the flattop period for cavities in ACC6 when both FB and AFF are off. It can be seen that the pulse-to-pulse signal shape varies smoothly along each pulse, suggesting that the changes are due to integrated effects as opposed to fast changes such as those caused by dark currents or multipacting. The PSD of the cavity probe signals at each sample point during the flattop period are plotted for all cavities in Figure 12. The jitter is typically a few tenths of percent at the beginning of flattop period and grows up to 4% by the end of flattop period.

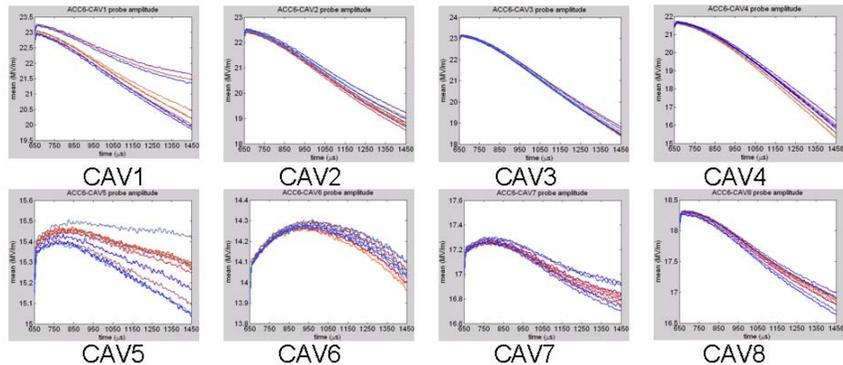

Figure 11: Selected cavity probe signals during the flattop period for cavities in ACC6.

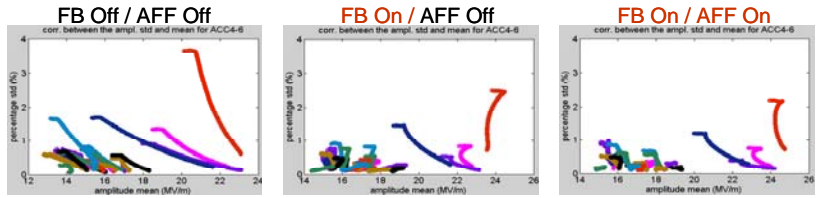

Figure 12: PSD of all cavity probe signals at each sample point during the flattop period.

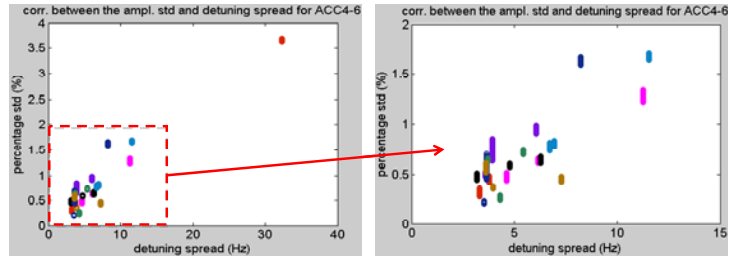

Figure 13: Correlation between cavity probe PSD and detuning ASD.

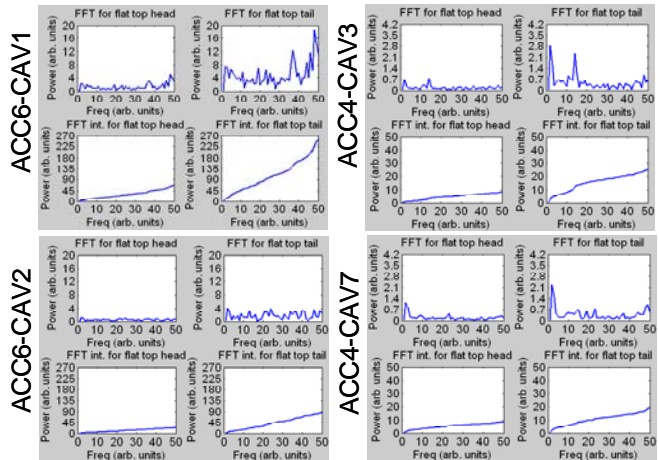

Figure 14: FFT and integrated FFT of the probe signals for four cavities at a sample time at the beginning (head) and end (tail) of the flattop period.

Figure 13 shows that there is a strong correlation between the cavity probe PSD and the detuning jitter at the end of the flattop period (measured by computing the cavity phase time derivative just after the input rf goes to zero). Thus variations in the pulse-to-pulse cavity detuning are likely driving the probe signal fractional jitter. Figure 14 shows Fast Fourier Transform (FFT) spectra of the cavity probe signals at the beginning and end of the cavity probe flattop period. Although there are some peaks in the FFT spectra, they do not contribute significantly to the integrated spectra (corresponding to the jitter ASD values in time domain). This indicates that the jitter is essentially random pulse-to-pulse. Also, the computed jitter correlation coefficients between cavities show that the cavity-to-cavity jitter is essentially uncorrelated.

In general, higher gradient cavities have higher jitter, but the jitter does not scale with the square of cavity gradient as would be expected in a simple Lorentz force detuning interpretation. As noted above, either the mechanical stiffness varies significantly among the cavities and/or some are vibrating more than others due to local external forces (in the case of the first cavity in ACC6, vibrations from the 4 K cooling system of the nearby quadrupole magnet is thought to be a contributing factor).

## 4    Conclusions and future plans

Overall, the LLRF system at the FLASH facility performs well in reducing the vector sum amplitude jitter to less than 0.1% and the phase jitter to less than $0.1^o$. The fractional jitter of the input rf is at the 0.1% level, which is excellent. The cavity probe signals are particularly interesting as their jitter is much larger (up to 4%) and grows along the pulse. The source of this jitter needs to be better understood in the future measurements as it may have important implications for XFEL and ILC.

There are number of effects that still need be evaluated, in particular, the reduction in the rf overhead afforded by piezo compensation, and the jitter that such compensation may introduce. Also, it would useful to measure how the probe signal jitter varies with gradient as higher gradients than those in this study are required for ILC. The plan is to continue to take more data for these cavities with/without piezo compensation and with beam on/off at different gradient and feedback gain levels. This time we will try to record the complex IQ data instead of the amplitude and phase data, and to record data at 5 Hz instead of 1/3 Hz.

## 5    Acknowledgements

The authors thank Nick Walker, Stefan Simrock, Valeri Ayvazyan, Zheqiao Geng (DESY), Gustavo Cancelo, Brain Chase (FNAL) and Shinichiro Michizono (KEK) for helpful suggestions and discussions. Also, we appreciate Michael Davidsaver's (FNAL) help with the data collection scripts.